\documentstyle [12pt,aasms4]{article} 

\begin{document}

\title{Physical Properties of Trans-Neptunian Object (20000) Varuna}  
\author{David C. Jewitt and Scott S. Sheppard}    
\affil{Institute for Astronomy, University of Hawaii, \\
2680 Woodlawn Drive, Honolulu, HI 96822 \\ jewitt@ifa.hawaii.edu, sheppard@ifa.hawaii.edu}

\begin{abstract}  

We present new time-resolved photometric observations of the bright
trans-Neptunian object (20000) Varuna and use them to study the
rotation period, shape, and color.  In observations from 2001 February
and April, we find a best-fit two-peaked lightcurve with period
$6.3442\pm 0.0002$ hr.  The peak-to-peak photometric range in the
R-band is $0.42\pm 0.02$ mag.  We find no rotational variation in
colors over the $0.45\leq \lambda \leq 0.85 \micron$ wavelength range.
From the short double-peaked period and large amplitude we suggest that
Varuna is an elongated, prolate body perhaps close in shape to one of
the Jacobi ellipsoids.  If so, the ratio of the axes projected into the
plane of the sky is 1.5:1 and the density is near $1000$ kg m$^{-3}$.
(20000) Varuna may be a rotationally distorted rubble pile, with a weak
internal constitution due to fracturing by past impacts.  The high
specific angular momentum implied by our observations and recent
detections of binary Trans-Neptunian Objects both point to an early,
intense collisional epoch in which large Trans-Neptunian Objects were
$\sim$100 times more abundant than now. In order to maintain a
cosmochemically plausible rock:ice mass ratio $\sim$ 0.5, Varuna must
be internally porous.

\end{abstract}


\section{Introduction}

The trans-Neptunian objects (TNOs) have semi-major axes larger than
that of Neptune (Jewitt and Luu 2000) and are thought to be the
products of arrested growth in the tenuous outer parts of the
accretion disk of the sun (Kenyon and Luu 1999).  Their large mean
heliocentric distances and resulting low surface temperatures suggest
that they may retain a substantial volatile fraction.  Indeed, the
trans-Neptunian region is widely held to be the source of the ice-rich
nuclei of short period (specifically Jupiter-family) comets.  In this
sense, the TNOs are repositories of some of the solar system's least
evolved, most primitive material.  There is widespread interest in the
physical (and chemical) properties of these bodies.  Unfortunately,
most known TNOs are too faint to permit easy investigation even when
using the largest available telescopes.  For this reason, the bright TNO
(20000) Varuna (apparent red magnitude $\sim$ 19.7) has already attracted
considerable observational attention.

Varuna was discovered on 2000 November 28 and given the provisional
designation 2000 WR106 (McMillan and Larsen 2000).  Prediscovery
observations from 1955 were soon uncovered (Knofel and Stoss 2000),
leading to the accurate determination of the orbit and the
classification as a "classical TNO" (Jewitt and Luu 2000), with
semimajor axis 43.274 AU, inclination $17.1\arcdeg$ and eccentricity
0.056.  Simultaneous thermal and optical observations give red
geometric albedo $p_{R}$ = 0.07$^{+0.030}/_{-0.017}$ and equivalent
circular diameter 900$^{+129}/_{-145}$ km (Jewitt, Aussel and Evans
2001).  Varuna is currently one of the biggest known Trans-Neptunian
objects and is comparable in size to the largest main-belt asteroid, 1
Ceres.  Farnham (2001) reported a rotational lightcurve from
observations taken 2001 Jan $24-27$ with a single-peaked period 3.17 hr
and "amplitude" 0.5 mag.  Farnham also found other plausible periods
including 2.78 and 3.67 hrs.  Motivated by this remarkable
result, we immediately undertook observations to secure an independent
determination of the lightcurve, and to search for rotational color
variations.  Such variations are predicted by the impact resurfacing
model (Luu and Jewitt 1996).  In this paper we will discuss Varuna's
lightcurve, possible causes of the brightness variations, and what this
object may reveal about the collisional environment in the 
young Trans-Neptunian belt.

\section{Observations}

Optical observations were taken UT 2001 February 17 - 21 and April 22,
24 and 25 at the University of Hawaii 2.2-m telescope atop Mauna Kea,
Hawaii.  We used a Tektronix $2048 \times 2048$ pixel charge-coupled
device (CCD), having 0.219 arcsec per pixel image scale and a 7
arcminute field of view.  Images were secured through standard
broadband BVRI filters based on the Johnson Kron-Cousins system.  The
CCD bias level was determined from an overclocked region of the chip.
Flat field calibration was obtained using a mixture of images of the
twilight sky with data frames.  The image quality (including
contributions from the telescope, wind shake and atmosphere) varied
from 0.6 to 1.2 arcsec Full Width at Half Maximum (FWHM), but was
mostly concentrated near a mean at 0.7 arcsec.  Absolute calibration of
the data was obtained through repeated observations of standard stars
from the list by Landolt (1992).  The position of Varuna was dithered
over the detector to prevent pathological problems in the photometry
associated with bad pixels.  Images photometrically affected by
proximity to bad pixels or field stars and galaxies (including all
those from UT 2001 February 21), were rejected from further analysis.

For the purposes of lightcurve determination, Varuna was compared to a
network of field reference stars in each image.  The field stars were
selected to be near Varuna in the sky and to have Sun-like or slightly
redder colors.  Where possible, we employed the same field reference
stars each night to minimize systematic photometric errors. This
procedure effectively removes the influence of atmospheric extinction
(which, at the high altitude of Mauna Kea, is already small, amounting
to only about 0.08 mag. per airmass in the R-band).  The latter part
of the night of UT 2001 February 19 was slightly non-photometric
(deviations were $\sim$ few 0.01 mag.) but, otherwise, we benefited
from clear skies.  Effects of seeing variations between images were
minimized by using multi-aperture photometry.  Small apertures
(typical radii $\sim$1 arcsec) were used to relate Varuna to the field
star network in each field.  Large apertures (typical radii $\sim$ 3.0 arcsec) were
used to relate the field stars to the Landolt standards.  The median
sky level was determined within a contiguous annulus having outer
radius 5 arcsec.  At the time of the observations, Varuna moved
westward at about 2 arcseconds per hour.  The image trailing due to
the motion of Varuna was only $\sim$0.1 arcseconds during a typical
200 second integration.  This is small compared to our nominal 0.7
arcsecond FWHM image quality and, therefore, trailing losses are an
unimportant source of error in our photometry.

A journal of observations including the geometric circumstances is
given in Table 1.  The R-band photometry is listed in Table 2.
Periodicity in the photometry was obvious even in preliminary
reductions of the first night's data conducted at the telescope, as
may be seen in Figure 1.

\section{Results}

The apparent magnitude varied in the approximate range 19.50 to 19.95,
with a mean near 19.7.  Periodicity in the R-band photometry was sought
using a) the PDM (phase dispersion minimization) method (Stellingwerf
1978) and b) the related but different SLM (string length minimization)
method (Dworetsky 1983).  The PDM and SLM results were in all cases
consistent; for brevity we here present only the results from PDM.
Figure 2 shows the PDM theta parameter as a function of rotational
frequency for the entire R-band photometry data set (the closer theta
is to zero, the better the fit: see Stellingwerf (1978) for more
information).  Broad, deep minima occur near the lightcurve frequencies
$P^{-1} = 7.57$ day$^{-1}$ and $P^{-1} = 3.78$ day$^{-1}$.  The first
corresponds to a single-peaked lightcurve with a period near $P = 3.17
$ hr, close to the 3.17 hr period reported by Farnham (2001). The
second minimum corresponds to the double-peaked lightcurve of $6.34$
hr.  The minima are flanked by aliases due to the 24-hour sampling
periodicity that is imposed on the data by the day-night cycle (see the
theta minima in Figure 2 displaced from the primary by $\pm n$
day$^{-1}$, where $n = 1$, 2, 3 ...).  The phased lightcurves produced
using the 24-hr alias periods of the single-peaked lightcurve ($P^{-1}
= 6.5$ day$^{-1}$, 8.6 day$^{-1}$) are unconvincing.  When viewed at
high resolution, the primary minimum of the theta plot for the
single-peaked lightcurve is seen to be split by a finely spaced series
of minima due to the $\sim$65 day data gap between the February 2001
and April 2001 observations (Figure 3).  The lightcurves produced by
phasing the data at each of the lowest 3 minima in Figure 3 (3.1656,
3.1721, and 3.1788 hours) appear comparably good to the eye.  In this
sense, the rotation period cannot be exactly determined from the
present data, although we are confident that, if Varuna's lightcurve is
single-peaked, the true period is given by one of the 3 minima in
Figure 3.  Subject to this caveat, we adopt $P = 3.1721 \pm 0.0001$ hrs
as the best fit to the single-peaked lightcurve period.

Is the lightcurve single or double-peaked?  Close inspection of the raw
and phased data suggests that the lightcurve of Varuna has two maxima
per rotation period.  This is evident in Figure 4, where phased
lightcurves show that the first and second lightcurve minima have
slightly different shapes.  Therefore, we adopt the double-peaked
lightcurve ($P_{rot}= 6.3442 \pm 0.0002$ hrs) as the probable rotation
period of Varuna (subject to the caveat that very nearby alias periods
6.3317 and 6.3574 hours are also plausible fits to the data as
discussed for the single peaked aliases above).  Our best estimate for
the photometric range of the data is $\Delta m = 0.42 \pm 0.02$ mag.

We used the preliminary photometry to target color measurements near
the extrema of the R-band brightness, in order to search for rotational
color variations.  Color measurements in B, V and I filters were
interleaved with R-band photometry so as to correct for photometric
trends caused by the rotation.  The color measurements are summarized
in Table 3 and plotted versus rotation phase in
Figure 5.  There it may be seen that our data provide no
evidence for rotational modulation of the color of the scattered
radiation.  Specifically, the B-V, V-R and R-I color indices are
constant with rotational phase at the level of accuracy of the
measurements, as seen in Table 3.  The mean colors found for Varuna,
$B-V = 0.85 \pm 0.02$, $V-R = 0.64 \pm 0.01$, $R-I = 0.62\pm 0.01$ and
$B-I = 2.11\pm 0.02$ (Table 3) are compatible with the mean colors of
12 other classical TNOs measured independently (Jewitt and Luu 2001:
$B-V = 1.00 \pm 0.04$, $V-R = 0.61 \pm 0.03$, $R-I = 0.60\pm 0.04$ and
$B-I = 2.22\pm 0.10$).  Although larger than most, Varuna is not colorimetrically
distinguished from the other TNOs.

\section{Interpretation}

The rotational lightcurve of a Solar System body results from the
combined effects of aspherical body shape and azimuthal albedo
variations.  Double-peaked lightcurves are the expected signature of a
prolate body in rotation about its minor axis.  However, this
interpretation is not unique: an appropriate arrangement of albedo
markings can reproduce any lightcurve (Russell 1906).  The large size
of Varuna suggests that any elongation of the body is probably caused
by a large specific angular momentum and the resulting rotational
deformation.  Here, we consider the two limiting cases in which a) the
lightcurve is produced entirely by albedo variations across the surface
of Varuna and b) the lightcurve is produced by rotational modulation of
the geometric cross section of Varuna due to aspherical shape.  The
real situation of Varuna will naturally lie somewhere between these two
extremes.

\subsection{Albedo Models}

A complex distribution of albedo markings could produce the observed
lightcurve.  If so, the $\Delta m_{R} = 0.42 \pm 0.02$ magnitude photometric
range would imply an albedo contrast $10^{0.4\Delta m_R} \approx 3:2$ or
greater (depending on the projection of the rotation vector into the
line of sight).  Some spherical outer Solar System bodies show large
albedo contrasts, notably Iapetus (Millis 1977) and Pluto (Buie, Tholen
and Wasserman 1997).  On Iapetus, the albedo
contrast is associated with a color variation, the dark material being
redder than the bright material (Table 4).  Rotational color variations
on Varuna as large as those on Iapetus would be apparent in our data if
they were present.  Pluto's hemispherical albedo contrast is matched by
a corresponding color variation that is barely measurable even in this
bright object (Table 4).  Large color differences exist between local
surface units on Pluto ($0.77 \leq B-V \leq 0.98$; Young, Binzel and
Crane 2001) but hemispherically averaged color variations occur only at
the 0.01 mag. level and are so small that they would not be detected in
the present work (Table 4).  From the example set by Pluto, we conclude
that the absence of color variations on Varuna larger than a few
$\times 0.01$ mag. places no useful constraint on the albedo
modulation hypothesis.

If Varuna is spherical and rotating at period $P$, we can
obtain a lower limit to the density by requiring that the body not be
in a state of internal tension.  Simple force balance then gives

\begin{equation}
\rho = \frac{3\pi}{GP^{2}}
\label{eq:albedoden}
\end{equation}

where $G = 6.67\times 10^{-11}$ [N kg$^{-2}$ m$^{2}$] is the
gravitational constant and $P [s]$ is the rotation period.  With $P =
3.17$ hr, Eq. (1) gives a lower limit to the density as $\rho = 1090$
kg m$^{-3}$.  The hydrostatic pressure at the center of a 450 km radius
object with this density is $\sim$ 6 $\times 10^7$ N m$^{-2}$.  A body
as large as Varuna will almost certainly not possess the strength
needed to maintain spherical shape when rotating with a 3 hr period
(note that the rotations of the comparison objects Iapetus and Pluto are
tidally locked, with periods 79 and 6.4 $\it days$, respectively, so
that rotational deformation is small).  Therefore, we believe that the
spherical albedo model of Varuna is physically implausible.  Instead,
we next discuss models in which the shape, not the albedo, is primarily
responsible for the observed lightcurve variations.

\subsection{Shape Models}

A more likely model is one in which Varuna is rotationally deformed by
the centripetal forces associated with its rapid rotation.  As in the
asteroid belt, large impacts that do not completely destroy a body may
disrupt it into a self-gravitationally bound, strengthless "rubble pile" (Farinella
et al. 1981).  Rubble-pile bodies will re-assemble after impact into a
shape determined by their angular momentum, $H$, and density, $\rho$.
Following the convention of Chandrasekhar (1987), we write the angular
momentum in units of $(GM^{3}a^{'})^{1/2}$, where $M$ [kg] is the body
mass and $a^{'}$ [m] is the radius of the equal-volume sphere.  At $H =
0$, the equilibrium shape is the sphere.  As $H$ increases, a perfectly
strengthless fluid deforms first into an oblate (so- called MacLaurin)
spheroid in rotation about its minor axis (Chandrasekhar 1987). The
MacLaurin spheroids become progressively more flattened up to the
critical value $H = 0.304$, above which the body becomes triaxial (a
Jacobi spheroid).  At $H \geq 0.390$, the Jacobi spheroids are
rotationally unstable and the object splits into a binary.  This
rotational deformation sequence is highly idealized when applied to
solid bodies, of course, because even a rubble pile will not respond to
rotational stresses in the same way as a perfectly strengthless fluid.
A fractured body will have a "grainy" structure, perhaps with large,
coherent internal blocks that will retain expression in the final body
shape.  Nevertheless, the MacLaurin-Jacobi spheroids represent a
limiting case with which the asteroids and TNOs may usefully be
compared.  Evidence for the existence of rubble-piles among the
main-belt asteroids is limited and indirect but nevertheless
suggestive.  For example, small asteroids ($D\leq 100$ km) have a
distribution of body shapes that is consistent with those of fragments
from hypervelocity impact experiments in the laboratory (Catullo et
al.  1984).  On the other hand, larger asteroids are deficient in
highly elongated bodies relative to the fragment shape distribution,
perhaps because their shapes have relaxed towards rotational
equilibrium (Farinella et al.  1981).  As we note below, a few well-
measured asteroids have densities less than the density of solid rock,
suggesting a porous internal structure that could be produced by
re-accumulation of large blocks in a rubble-pile (Yeomans et al.
1999).  TNOs larger than about 100 km in diameter are massive enough to
survive collisional disruption over the age of the solar system but may
nevertheless have been internally fractured into rubble piles
(Farinella and Davis 1996).

Since an oblate spheroid in principal axis rotation about its axis of
maximum moment of inertia offers no rotational modulation of the
cross-section, we conclude that the MacLaurin spheroids cannot explain
the lightcurve of Varuna, and that $H > 0.304$ for this object.
However, the triaxial Jacobi spheroids with $0.304\leq H\leq 0.390$
present plausible solutions for the shape of Varuna.  For the Jacobi
spheroids, knowledge of the rotation period and the shape (from $\Delta
m$) provide a unique measure of the density.  With $P$ = 6.34 hr and
axis ratio $10^{0.4\Delta m} \approx 3:2$, we obtain $\rho = 1050$ kg
m$^{-3}$ from the tables of Chandrasekhar (1987).  Since the axis ratio
can only be a lower limit (because of projection), the derived density
is also a lower limit.  This value is close to the density estimated
from the spherical albedo model, above, but has the advantage that it
is derived from a physically more plausible model.  A Jacobi
ellipsoid's three axes ($a>b>c$) depend strongly on each other.  By
using the axis ratio for $a$ and $b$ above we find the ratio for all
three as 3:2:1.4 (Chandrasekhar 1987). Here 1.4 refers to the rotation
axis ($c$) which can not be observed directly through the lightcurve if
the object is in principal axis rotation.  

\subsection{Binary}

We also consider the possibility that Varuna is a binary, in which
case the lightcurve would result from occultation of one component by
the other.  The short period and lightcurve shape of Varuna require
that the binary components must be close or even in contact, leading
to mutual gravitational deformation of the components.  We will
consider these effects momentarily.  However, it is physically
illuminating to first discuss the limiting case in which the binary
components are in contact but retain a spherical shape.  Suppose that the primary and
secondary components have radii $a_{p}$ and $a_{s}$, respectively.
The barycenter of the contact binary will be separated from the center
of the secondary by a distance

\begin{equation}
l = (a_{p}+a_{s})\left(\frac{m_{p}}{m_{p}+m_{s}}\right)
\label{eq:binsep}
\end{equation}

where $m_{p}$ and $m_{s}$ are the masses of the primary and secondary,
respectively.  Force balance then gives a relation between the
angular frequency, density and mass ratio, namely
$l\omega^{2}=Gm_{p}/(a_{p}+a_{s})^{2}$.  Assuming that both
components have density $\rho$ [kg m$^{-3}$], we obtain

\begin{equation}
\rho = \frac{3\pi}{GP^{2}}\left[\frac{(1+(a_{s}/a_{p}))^{3}}{1+(a_{s}/a_{p})^{3}}\right]
\label{eq:binrho}
\end{equation}

The ratio $a_{s}/a_{p}$ can be estimated from the photometric range of the
lightcurve.  Specifically, the range in magnitudes is given by

\begin{equation}
\Delta m = 2.5 \mbox{log}\left[\frac{a_{s}^{2}+a_{p}^{2}}{a_{p}^{2}}\right]
\label{eq:binmag}
\end{equation}

which can be rearranged to yield

\begin{equation}
\frac{a_{s}}{a_{p}} = (10^{0.4\Delta m}-1)^{1/2}
\label{eq:binrad}
\end{equation}

From Equation 5, the Varuna lightcurve range ($\Delta m = 0.42$ mag.)  suggests an axis
ratio $a_{s}/a_{p} = 0.69$, corresponding
to mass ratio $m_{s}/m_{p} = (a_{s}/a_{p})^{3} = 0.32$.
Equations~\ref{eq:binrho} and \ref{eq:binrad} together give the density in
terms of the two observable quantities $P$
and $\Delta m$:

\begin{equation}
\rho = \frac{3\pi}{GP^{2}}\left[\frac{(1+(10^{0.4\Delta m}-1)^{1/2})^{3}}{1+(10^{0.4\Delta m}-1)^{3/2}}\right].
\label{eq:binrhofinal}
\end{equation}

Substitution of $P = 6.3$ hrs, $\Delta m = 0.42$ mag.
into Equation~\ref{eq:binrhofinal} gives $\rho = 996$ kg m$^{-3}$.
This is a lower limit to the density since, due to projection effects,
the observations provide only a lower limit to $\Delta m$.  However,
given the crudeness of this model, it is interesting that we obtain
essentially the same density as from the Jacobi spheroid
approximation.   We conclude that binary models with the observed
period and physically plausible densities can account for the Varuna
lightcurve.  There remains, however, no specific evidence that Varuna
is a binary object.

Lastly, Leone et al. (1984) present calculations that partially account
for the mutual deformation of the components in a close binary.  The
solutions are non-unique, since different combinations of $m_{s}/m_{p}$
and density lead to the same $P$, $\Delta m$ pair.  Here, we simply
refer to their Table 1, and note a plausible solution near $\rho
\approx 3600$ kg m$^{-3}$ and $m_{s}/m_{p} \approx 0.2$.  This bulk
density is much higher than is required than in our simple, spherical
contact binary model since the components are more widely separated
(the Roche radius is $\sim$ twice the component radius) and
therefore orbit more slowly for a given mass.  For several reasons we
believe the Roche model and the Leone et al. calculations are not
well-suited to the case of Varuna.  In particular, the wide separation
reduces the probability that the system would be aligned so as to
produce mutual eclipses.  Furthermore, the nearly sinusoidal lightcurve
shape is not easily produced by a wide pair in which we would expect
the eclipses and occultations to occupy a smaller fraction of
rotational phase space.  Varuna's lightcurve does not show the notched
appearance characteristic of eclipsing binary stars.

\section{Discussion}
\subsection{The Density of Varuna}

The bulk density derived in Section 4 is only slightly higher than the
density of water ice.   Strictly speaking, the derived density is a
lower limit because of the effects of projection into the plane of the
sky.  However, it is unlikely that the density is much (factor of two)
higher than $1000$ kg m$^{-3}$ if our equilibrium rotator model is
correct.  The simplest interpretation, namely that the rock content of
Varuna is small or negligible, is difficult to accept on physical
grounds: accretion of planetary solids in a dusty circumsolar disk
provides no obvious fractionation mechanism to lead to the formation of
a pure ice ball.  Here, we explore the possibility that Varuna's low
density is caused by an internal structure that is at least partly
porous.

We represent Varuna as a composite of volatile matter (ice), refractory
matter (rock) and void space.  The mean density of this composite is

\begin{equation}
\overline{\rho} = \rho_if_i + \rho_rf_r
\end{equation}

where $\rho_i$ and $\rho_r$ are the densities of ice and rock, and
$f_i$ and $f_r$ are the fractional volumes occupied by ice and rock, respectively.
We require 

\begin{equation}
f_i + f_r + f_v \equiv 1,
\end{equation}

where $f_v$ is the fractional void
space, also known as porosity.  The fraction of the total mass carried by refractories is

\begin{equation}
\psi = \frac{\rho_rf_r}{\rho_rf_r + \rho_if_i}.
\end{equation}

Equations (7) - (9) combine to give 

\begin{equation}
f_v = 1 - \frac{\overline{\rho}}{\rho_i}\left[1 + \psi\left(\frac{\rho_i}{\rho_r}-1\right)\right].
\end{equation}

The hydrostatic pressure at the core of Varuna is about 6 $\times 10^7$
N m$^{-2}$, while the mass-weighted internal temperatures are likely to
average $\sim$ 100 K ($\sim$ 50 K at the surface, slowly rising towards
the core as a result of internal radioactive decay heating).  Under
these conditions the ice-I polymorph of water, for which $\rho_i$ = 940
kg m$^{-3}$ (Lupo and Lewis 1979), is the stable phase.  The density of
the refractory matter, $\rho_r$, is less certain and we compute models
using two representative values.  First, we take $\rho_r$ = 3000 kg
m$^{-3}$ as representative of the densities of a number of plausible
silicates, notably Forsterite ($Mg_2SiO_4$) which has been
spectroscopically detected in comets.  Second, we take $\rho_r$ = 2000
kg m$^{-3}$ to represent the less dense CHON type hydrocarbon materials
that are also found in comets.

The resulting values of the porosity are plotted against the rock
fraction in Figure 6, where we have adopted $\overline{\rho}$ = 1000 kg
m$^{-3}$ (solid lines) from the lightcurve models.  To show the
sensitivity to the mean density we also plot in Figure 6 the porosities
derived if $\overline{\rho}$ = 1200 kg m$^{-3}$ (dashed lines).

Figure 6 shows that non-porous ($f_v$ = 0) models restrict the rock
fraction to $\psi \sim 0.1$ ($\overline{\rho}$ = 1000 kg m$^{-3}$) to
$\psi \sim 0.35$ ($\overline{\rho}$ = 1200 kg m$^{-3}$).  Larger rock
fractions require non-zero porosity; cosmochemically plausible $\psi$ =
0.5 implies $0.05 \leq f_v \leq 0.30$ for the parameters considered
here.  The few measured nuclei of comets have densities in the range $500
- 1000$ kg m$^{-3}$ (Sagdeev et al. 1988, Rickman 1989, Asphaug and
Benz 1996) and rock fractions $\psi \geq$ 0.5 (corresponding to
rock/ice mass ratios $\geq$ 1, Lisse et al. 1998, Jewitt and Matthews
1999).  Given that the Jupiter family comets are collisionally produced fragments of
the trans-Neptunian objects (Farinella and Davis 1996), large $\psi$
may well be representative of Varuna and related objects.  

Are porosities up to several 10's of \% physically plausible in a body
of Varuna's size?  We argue first by analogy.  Materials on planetary
surfaces are commonly porous.  Carbonate ($CaCO_3)$ Waikiki beach sand,
for instance, has porosity $f_v \sim$ 0.4 while the basaltic lunar
regolith has $0.4 \leq f_v \leq 0.7$ (Carrier et al.  1974).  The
mechanical compression of carbonate and quartz sands ($SiO_2$) has been
studied experimentally to pressures rivalling that in Varuna's core (Chuhan et al.
2002). Sand
samples having porosity $f_v$ = 0.4 to 0.5 at atmospheric pressure
compress only to $f_v \sim$ 0.2 to 0.3 at 5 $\times 10^7$ N m$^{-2}$.  At
pressures $\leq$ (2 to 6) $\times 10^6$ N m$^{-2}$ the densification
occurs by rearrangement of the irregular silicate grains while, at
higher pressures, the densification occurs by grain crushing and the
filling-in of void space. Crushing is particularly important in
large-grain sands where the number of grain-grain contacts is small and
the contact pressures are high.  In any event, this direct experimental
evidence shows that aggregates of silicate (and carbonate) grains can
remain substantially porous throughout the range of pressures
prevailing inside Varuna.  The physical reason is the large compressive
strength of the grain materials.

When at low temperatures, water ice also possesses a large compressive
strength (1.2 $\times 10^8$ N m$^{-2}$ at 158 K rising to 1.7 $\times
10^8$ N m$^{-2}$ at 77 K; Durham et al.  1983) giving a response to
hydrostatic loading qualitatively similar to that of sand.  However,
the experimental situation for ice grain aggregates is less clear,
mainly because most experiments have been performed at temperatures and
strain-rates higher than are relevant to outer solar system bodies.
The most relevant experiments of which we are aware were conducted at
213 K over the 0.8 to 8.2 $\times 10^8$ N m$^{-2}$ pressure range
(Leliwa-Kopystynski et al. 1994).  They show that substantial porosity
can exist in ice/rock grain mixtures at all pressure levels found
inside Varuna.  We conclude that porosity is a potentially important
factor in determining the bulk constitutions of the TNOs.

Several of the outer planet satellites are similar to Varuna in both
size and density.  For example, Saturnian satellite SIII Tethys has
density $\overline\rho = 1210\pm 160$ kg m$^{-3}$ and is $524\pm 5$ km in radius
(Smith et al. 1982).  The Uranian satellite UII Umbriel has density
$1440\pm 280$ kg m$^{-3}$ and radius $595\pm 10$ km (Smith et al.
1986).  Even the much larger and, presumably, self-compressed object
Iapetus (radius $730\pm 10$ km) has a low density of only $1160\pm 90$
kg m$^{-3}$ (Smith et al.  1982).  Internal porosity (due to the
granular structure of the constituent materials) may account for the
low densities of these satellites while simultaneously allowing rock
fractions 0.28 $\leq \psi \leq$ 0.66 (Kossacki and Leliwa-Kopystynski
1993).  Within the uncertainties, these bodies all have densities
consistent with that derived here for Varuna.  Unlike Varuna, they are
nearly spherical in shape, but this is because the satellites are
tidally locked with rotation periods measured in days, not hours, and the
centripetal accelerations are consequently very small.  If rotating
with Varuna's angular momentum, they would adopt prolate body
shapes and display large rotational lightcurves.  

The larger object Pluto has $\rho \approx 2000$ kg m$^{-3}$ (Tholen and
Buie 1997): the corresponding rock-fraction from Eq. (10) (assuming
$f_v$ = 0 and considering higher pressure ice phases) is $\psi \sim$
0.7.  Core hydrostatic pressure (which scales as $\rho^2r^2$) is larger
in Pluto than in Varuna by a factor $\sim$25.  For this reason it is
natural to expect that the influence of porosity on the bulk density
should be greatly reduced relative to the Varuna case.

A different source of porosity is suggested by the 50 km diameter and
(presumably) rocky main-belt asteroid 253 Mathilde, which is of
surprisingly low density ($1300\pm 200$ kg m$^{-3}$; Yeomans et al.
1999).  This may reflect macroscopic porosity ($\sim$ 50\%) resulting
from the loose re-accumulation of a fractured body into a rubble-pile
structure.  In Varuna, microscopic and macroscopic porosity may
co-exist, the former from the slow accretion of grains at low temperatures
and the latter from an energetic, terminal collision phase.

Varuna is one of the largest known trans-Neptunians, with a diameter $D
= 900(+125/-145)$ km ($1\sigma$ error bars; Jewitt, Aussel and Evans
2001).  Even the 3$\sigma$ lower limit to the diameter, 465 km, leaves
this as an object of imposing dimensions.  But Varuna is not entirely
alone in its combination of large size, short period and aspherical
shape: there are a few large asteroids which have both large
photometric ranges and short periods.  Notable examples include 15
Eunomia (diameter 256 km, period 6.08 hr, photometric range 0.56 mag.),
87 Sylvia (270 km, 5.18 hr, 0.62 mag.) and 107 Camilla (222 km, 4.84
hr, 0.52 mag.).  Like Varuna, these may be bodies of low strength which
are rotationally deformed due to a large amount of angular momentum
delivered by collisions (Farinella et al. 1981).  Collectively,
these observations show that the low density of Varuna, as deduced from
its lightcurve, is unremarkable when viewed in the context of other
solar system objects.

\subsection{High Specific Angular Momentum in the Trans-Neptunian Belt}

The discussion in Section 4 shows that Varuna must possess a high specific
angular momentum, $0.304 \leq H \leq 0.390$, close to the value needed
to cause rotational breakup.  This immediately suggests a parallel with
the 4 known Trans-Neptunian binaries, Pluto-Charon (Tholen and Buie
1997), 1998 WW31 (Veillet 2001), 2001 QT297 (Elliot et al.  2001) and
2001 QW322 (Kavelaars et al. 2001) because these systems also possess
high specific angular momenta.  Of these four, Pluto-Charon is by far
the best characterised (neither the sizes/masses nor the eccentricities
of the other binary systems are yet known, preventing the determination
of their specific angular momenta at a diagnostically useful level).
In the Pluto-Charon system the angular momentum is primarily contained
within the orbital motion of the components (i.e., the spin angular
momentum is small).  If the mass and angular momentum of Pluto and
Charon were to be combined into a single body, that object would have
$H \approx 0.45$ (McKinnon 1989) and would be unstable to rotational
breakup.  This fact suggests that Pluto and Charon were formed by a
glancing impact between precursor objects (McKinnon 1989, Dobrovolskis
et al. 1997).  It is not yet clear that the other (much less massive)
TNO binaries formed in a similar way, but their high specific angular
momenta nevertheless suggest that collisions played an important role.
We believe that Varuna is an intermediate case in which a collision
between massive precursors produced a spin rapid enough to cause global
deformation but not sufficient to cause rotational breakup.

The current collision rate amongst Trans-Neptunian Objects is too low
to substantially modify the spins of objects as large as Pluto and
Varuna.  Instead, their $H$ must have been acquired through late-stage
collisions at the end of the $\sim$ 100 Myr (Kenyon and Luu 1999)
formation period.  We use this to constrain the collisional environment
in the young Trans-Neptunian belt.  Since the applicable parameters are
not well known, an order of magnitude (particle-in-a-box) type
calculation is appropriate.  Our objective is to estimate the density
of objects in the Trans-Neptunian belt needed to produce collisional
breakup of a measurable fraction of the Varuna-sized objects within the
100 Myr formation period.  Collisional breakup does not guarantee the
formation of a binary, and may not lead to a high specific angular
momentum in the target object, but it is a necessary step towards these
ends.

We first represent the TNO size distribution by a power law, in which
the number of objects per unit volume with radii in the range $a$ to $a
+ da$ is $n(a)da$ = $\Gamma a^{-q} da$, where $\Gamma$ and $q$ are
constants.  The timescale for the collisional disruption of a
non-rotating, spherical target TNO of radius $a_T$ is given by

\begin{equation}
\tau^{-1} = \int_{a_C}^{\infty} 4 \pi \left(a_T + a\right)^2 \Gamma a^{-q} \Delta V da.
\end{equation}

Here, $a_C$ is the critical radius of the smallest projectile capable
of producing disruption and $\Delta V$ is the impact velocity.  The
constant $\Gamma$ is obtained from the data by normalizing to the
Trans-Neptunian belt population, such that

\begin{equation}
\int_{a_{50}}^{\infty}\Gamma a^{-q} da = \frac{N_{50}}{W}
\end{equation}

where $N_{50}$ is the number of TNOs larger than $a_{50}$ = 50 km in radius,
and $W$ is the effective volume swept by the classical TNOs.  We take
$q$ = 4, $\Delta V$ = 1.3 km $s^{-1}$, $N_{50}$ = 4 $\times 10^4$
(Trujillo, Jewitt and Luu 2001) and represent the classical belt as an
annulus with inner and outer radii of 40 AU and 50 AU, respectively,
and thickness 10 AU, giving $W \approx 9.5 \times 10^{38}$ $m^3$.
Hence, $\Gamma$ = 1.6 $\times 10^{-20}$ by Eq. (12).

For spherical target and projectile TNOs of
equal density, the critical projectile radius for disruption is given by

\begin{equation}
\frac{a_C}{a_T} = \left(\frac{2 \epsilon}{\Delta V^2}\right)^{1/3}.
\end{equation}

where $\epsilon \approx 10^5$ J kg$^{-1}$ is the
specific energy for disruption (Love and Ahrens 1996).  

Combining these relations and substituting, we obtain 

\begin{equation}
\tau \approx 2.5 Gyr \left[\frac{a_T}{1 km}\right] \left[\frac{4 \times 10^4}{N_{50}}\right]
\end{equation}

as our estimate of the collisional lifetime of a TNO of radius $a_T$
in the present day, low-density Trans-Neptunian belt. Varuna-scale objects ($a_T
\sim 450$ km) have lifetimes to collisional disruption $\tau \sim 1
\times 10^{12}$ yr (Eq.  14), much longer than the 4.6 $\times 10^{9}$
yr age of the solar system (Farinella and Davis 1996).  Under these
circumstances, large collisionally produced binaries and objects of high
specific angular momentum should not be common. 

The frequencies of occurence of rotationally distorted and binary TNOs
both remain to be determined with accuracy.  Objects 1998 SM165 and
2000 GN171 have independently been reported to show large rotational
variations (Romanishin et. al 2001, Sheppard 2001) and, from the
available data, it appears likely that large amplitude objects like
Varuna are common, as are binary TNOs.  We conservatively estimate that
a fraction $f \geq$ 1\% of the TNOs are binaries and/or rotationally
distorted objects of high intrinsic angular momentum. If the formation
phase took $\tau_f \sim 10^8$ yr (Kenyon and Luu 1999), then the corresponding
disruption timescale is of order $f^{-1} \tau_f \leq$ $10^{10}$ yrs.
With this timescale and $a_T$ = 450 km (for Varuna), Eq. (14) gives
$N_{50} \sim$ 5 $\times 10^6$, about 100 times the current number of
large TNOs.  The current mass of the TNOs is about 0.1 to 0.2
$M_{\oplus}$ (Jewitt, Luu and Trujillo 1998, Trujillo, Jewitt and Luu
2000).  Thus, from the high incidence of binaries and rotationally
distorted objects, we infer an initial mass $M_i \geq$ 10 to 20
$M_{\oplus}$ in the Classical region of the Trans-Neptunian belt.
Numerical accretion calculations give the time for Trans-Neptunian
objects to grow to 1000 km scale as $\tau \sim$ 280 Myr (10 $M_E$ /
$M_i$) (Kenyon and Luu 1999).  Substituting for $M_i$ we obtain $\tau
\sim $ 140 to 280 Myr, comparable to the $\sim$ 100 Myr timescale
estimated for the formation of Neptune.  

Efficient growth of the TNOs required a small velocity dispersion.
Growth models show that mutual scattering by ever larger bodies
produced an increase in the velocity dispersion, ultimately resulting
in collisional shattering of the smaller objects (Kenyon and Luu
1999).  The broad inclination and eccentricity distributions of the
current TNOs suggest that another process, perhaps the appearance of
nearby Neptune or, conceivably, an external perturber (Ida et al. 2000),
increased the velocity dispersion to values even higher than could be
attained through mutual scattering.  The collisional events leading to
Varuna's rapid rotation and the break-up of TNOs probably occured in
this final stage.

On-going measurements of the lightcurves, spin states and binarity of a
statistical sample of TNOs will show the extent to which the shapes of
these bodies are determined by their angular momenta (see Sheppard and
Jewitt 2002).  Detection of a large proportion of high $H$ objects would
strengthen our conclusion that the past environment in the Trans-Neptunian
belt was one in which collisions between massive bodies were common.
The role of porosity will be illuminated by future measurements of TNO
densities, perhaps based on careful astrometric and physical studies of
the growing sample of binary objects.  We expect a trend towards higher
densities at larger diameters, at least amongst those TNOs that have
survived intact (although fractured) since the formation of the solar
system.  Smaller objects, particularly the 1 to 10 km scale bodies
sampled locally in the Jupiter-family comets may display a wide range
of densities influenced both by the density (hence, size) of the target
from which they were eroded and by densification during the collisional
process.

\section{Summary}

1) High precision, time series photometry of trans-Neptunian object
(20000) Varuna reveals a sustained, periodic modulation of the apparent
brightness.  In data from 2001 February and April, the best-fit
rotation period is double-peaked with a $6.3442 \pm 0.0002$ hour
period.  The photometric range is $\Delta m = 0.42 \pm 0.02$ mag.
The B-V, V-R and R-I colors are completely typical of TNOs and show no
variation with rotational phase, down to the level of a few per cent.

2) We interpret Varuna as a centripetally distorted, prolate ellipsoid, with
an axis ratio $\geq$ 3:2 and density $\rho \approx 1000$ kg m$^{-3}$.  The
specific angular momentum, $0.304 \leq H \leq 0.390$, is high (objects with
$H$ $>$ 0.390 are rotationally unstable and break up).

3) The low bulk density of Varuna requires significant porosity (up to
several 10's of \%) if the rock mass fraction is cosmochemically
typical ($\sim 0.5$).  In this respect, Varuna is similar to the icy
satellites of Saturn and Uranus, many of which show bulk densities near
that of water ice.  Porosity could occur on a microscopic level due to
granular structure of the constituent material, or macroscopically if
Varuna has been loosely re-assembled after fracturing by past
impacts.

4) The high specific angular momentum of Varuna cannot have been supplied
by collisions in the present-day, low-density Trans-Neptunian belt.  High $H$ leading
to rotational deformation and binary formation suggests an early, high
density phase in the Trans-Neptunian belt.  We infer an initial mass $M_i \sim$
10 to 20 $M_{\oplus}$ from the apparently high incidence of binaries and rapidly
rotating Trans-Neptunian Objects.

\section*{Acknowledgments}

We thank John Dvorak and Paul deGrood for operating the UH telescope,
Jane Luu, Yan Fernandez and the anonymous referee for comments on the
manuscript and Fawad Chuhan for pointers about the literature of sand.
This work was supported by a grant to DCJ from the National Science
Foundation.

\newpage

\begin{figure}
\caption{Sample rotational lightcurve data for Varuna from UT 2001
February 17.  Error bars of ±0.02 mag. have been plotted for
reference.} \label{fig:nt2rot}
\end{figure}

\begin{figure}
\caption{Phase dispersion minimization (PDM) plot computed from the
entire R-band data set of Varuna (February 2001 and April 2001
observations).  The best fit is the frequency near 3.78 cycles per day
(double-peaked period of 6.34 hours).  The other large peaks flanking
the 3.78 frequency are the 24 hour sampling aliases.  The
single-peaked period is at 7.57 cycles per day (period of 3.17 hours)
with associated flanking 24 hour alias periods.}
\label{fig:pdmbig}
\end{figure}

\begin{figure}
\caption{Same as Figure 2 but plotted with the Period as the x-axis
and at higher resolution on the single-peaked best fit to show the
aliases caused by the $\sim$60 day gap between the February 2001 and
April 2001 observations.  The 3 lower peaks are all reasonable fits to
the data, with the middle peak at 3.1721 hours being the best fit for
the single-peaked lightcurve.}
\label{fig:pdmsmall}
\end{figure}

\begin{figure}
\caption{The R-band photometry of Varuna phased according to the
double-peaked rotation period $P_{rot} = 6.3442$ hours.  The April
data has been brightened by 0.09 magnitudes to correct for the dimming
effects of a higher phase angle (see Sheppard and Jewitt 2002) and
further distance of Varuna compared to the February observations.}
\label{fig:phased}
\end{figure}

\begin{figure}
\caption{The B-V, R-I, and V-R colors of Varuna showing no significant
variation over its rotation.  The R-I colors have been shifted up by
0.5 magnitudes for clarity in distinguishing them from the V-R
plotted colors.}
\label{fig:color} 
\end{figure}

\begin{figure}
\caption{Porosity, $f_v$, versus rock fraction, $\psi$, from Eq. (10).  Models for
$\overline\rho$ = 1000 kg m$^{-3}$ and $\overline\rho$ = 1200 kg
m$^{-3}$ are indicated.  The upper(lower) line for each density refers
to an assumed rock density $\rho_r$ = 3000(2000) kg m$^{-3}$.  The range
of rock fractions inferred for the icy satellites of Saturn and Uranus
by Kossacki and Leliwa-Kopystynski (1993) is shown for comparison.  }
\label{fig:porosity} \end{figure}

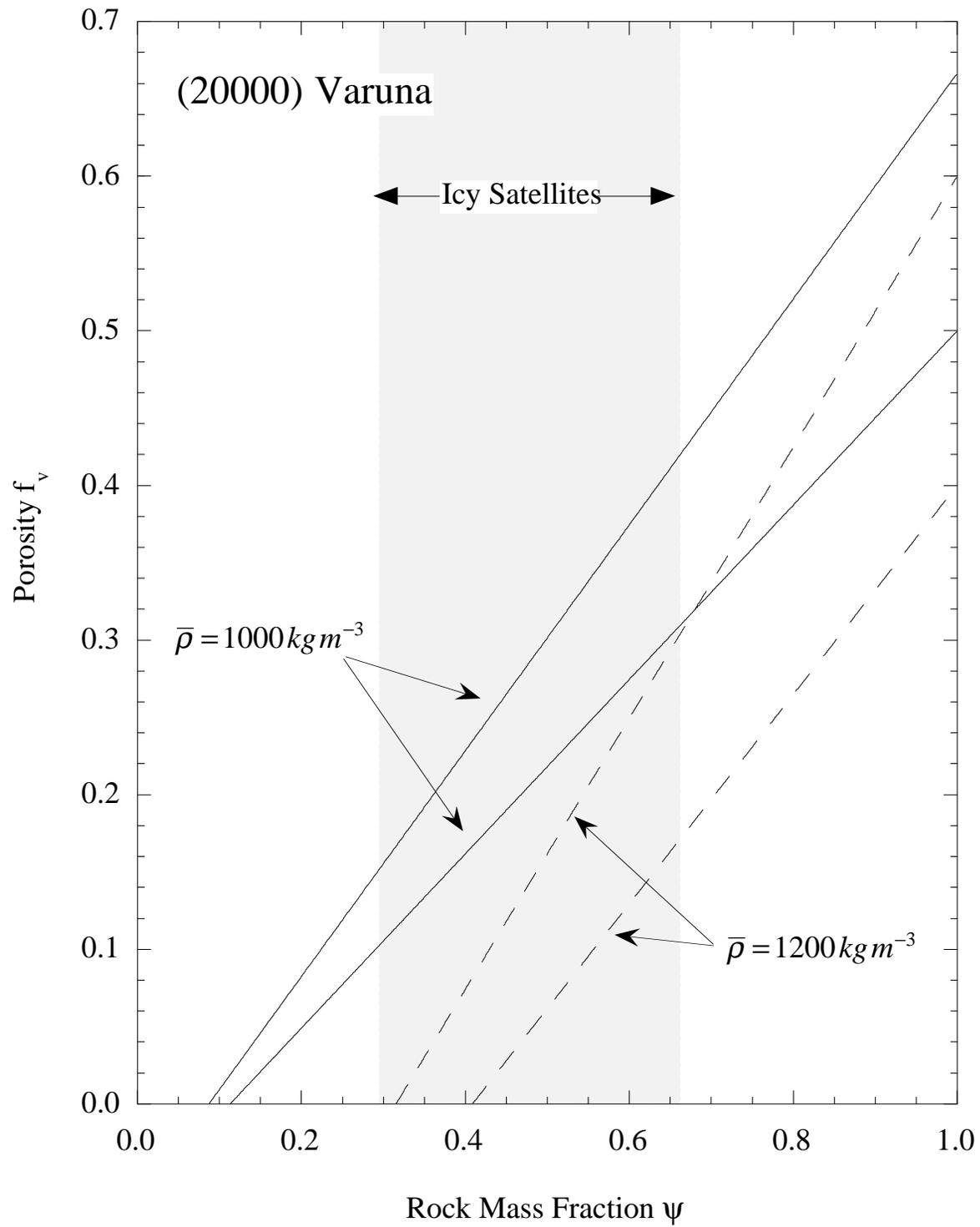
\end{document}